\documentclass[twocolumn,letter]{IEEEtran}
\usepackage[usenames,dvipsnames]{color}
\usepackage{subfigure}
\usepackage{graphicx}
\usepackage{amsmath}
\usepackage{color}
\usepackage{multicol}
\usepackage{amssymb}

\usepackage{balance}
\usepackage{amsfonts}%
\usepackage{verbatim}%
\usepackage{enumerate}%
\usepackage{psfrag}
\usepackage{epsfig}
\usepackage{multicol}
\usepackage{setspace}
\usepackage{dsfont}

\usepackage{algorithm}
\usepackage{algorithmic}
\usepackage{cite}
\usepackage{float}
\usepackage{cite}
 
\begin{document}
\title{Spectrum--Aggregating Cognitive Multi-Antenna User with Multiple Primary Users}
\author{ Ahmed El Shafie$^\dagger$, Tamer Khattab$^*$\\
\small \begin{tabular}{c}
$^\dagger$Wireless Intelligent Networks Center (WINC), Nile University, Giza, Egypt. \\
$^*$Electrical Engineering, Qatar University, Doha, Qatar.
\end{tabular}
}
\date{}
\maketitle
\thispagestyle{empty}
\pagestyle{empty}
\begin{abstract}
We investigate a cognitive radio scenario involving a single cognitive transmitter equipped with $\mathcal{K}$ antennas sharing the spectrum with $\mathcal{M}$ primary users (PUs) transmitting over orthogonal bands. Each terminal has a queue to store its incoming traffic. We propose a novel protocol where the cognitive user transmits its packet over a channel formed by the aggregate of the inactive primary bands. We study the impact of the number of PUs, sensing errors, and the number of antennas on the maximum secondary stable throughput.
\end{abstract}
%%%\vspace{-0.5 cm}
\begin{IEEEkeywords}
%%\vspace{-0.3 cm}
Cognitive radio, queue stability, multiple access, antennas, sensing errors, dominant system.
\end{IEEEkeywords}

%%\vspace{-0.40cm}
\section{Introduction}
%%\vspace{-0.20cm}
One challenge in cognitive radio networks is designing an optimal channel access for the secondary users (SUs) under certain quality of service for the primary users (PUs). This opens the research area for the invention of novel medium access protocols that enhance the performance of the network.

 Considering a single cognitive transmitter-receiver pair in presence of multiple primary transmitter-receiver pairs has got extensive attention in the literature \cite{zhao2007decentralized,zhao2008opportunistic,jsac,2013}. The authors of \cite{zhao2007decentralized} developed the optimal policy under the finite-horizon partially
observable Markov decision processes (POMDPs) formulation
that has a complexity growing exponentially with the duration
of the transmission. In that work, the overall state of the network is
partially observable due to the fact that a secondary transmitter senses only some of
the available channels. Under the assumption of having the same transmission structure for both primary and secondary
users, the authors derived the optimal and suboptimal spectrum sensing and
access strategies under the formulation of finite-horizon POMDPs.

 In \cite{zhao2008opportunistic}, the authors considered an infinite-horizon
optimization where the complexity does not grow with the
length of the transmission in contrast to \cite{zhao2007decentralized}. The authors assumed that the multiple PU channels evolve independently as
continuous-time Markov chains. Based on such assumption, the authors proposed an access scheme
referred to as periodic sensing opportunistic spectrum access
(PS-OSA). The essence of PS-OSA is to remove the partial observability
by sensing the available channels periodically. In general, restricting to periodic sensing is suboptimal, but the proposed
scheme significantly reduces the complexity required by
the optimal opportunistic spectrum access (OSA) proposed in \cite{zhao2007decentralized} under the POMDP framework.
The authors showed that
when the constraints on interference are tight, the performance
loss of PS-OSA is negligible.

 In \cite{jsac}, the PUs are modeled as independent continuous-time Markovian on-off processes. The secondary transmitter aims at maximizing its throughput subject to collision constraints. The authors investigated some access policies for the SU: sensing one primary channel, sensing all primary channels simultaneously, and memoryless with periodic sensing. The case of multiple SUs is also discussed. In \cite{2013}, the authors designed an OSA in the presence of reactive PUs, where PU's access probability in a given channel is related to SU's past access
decisions. The channel occupancy of the reactive PU is modeled as a four state discrete-time Markov chain and the optimal OSA design for SU throughput maximization is formulated as a
constrained finite-horizon POMDP problem.

In a cognitive setting with buffered nodes, the authors in \cite{bao2010stable} considered multiple PUs with a common destination and one cognitive radio user with relaying capability. When all primary packets being served in all primary and relaying queues, the cognitive radio user switches to the best idle band which has the maximum channel gain, based on the channel conditions, for the transmission of its own packets.
In \cite{krikidis2010stability}, Krikidis {\it et al.} considered a simple configuration composed of one cognitive transmitter-receiver pair and two primary transmitter-receiver pairs wishing to deliver their packets to a single destination in a multi-access
channel (MAC). The secondary transmitter is capable of relaying the undelivered packets of the PUs. If a primary packet is correctly decoded at either the secondary transmitter or the primary destination, it is dropped from the relevant primary queue. A priority of transmission is given to the relaying packets over the secondary own packets when the primary queues are empty. In addition, the secondary transmitter sends its own packets in two ways: 1) when all the primary and relaying queues are empty or 2) simultaneously with the PUs via a superposition technique when the primary queues are nonempty.

In this work, we investigate a cognitive scenario with one cognitive radio user (SU) possessing  $\mathcal{K}$ antennas and $\mathcal{M}$ single antenna PUs operating on orthogonal channels. The SU is capable of sensing and aggregating the idle channels to increase its transmission reliability.

The contributions of this paper can be summarized as follows. A new cognitive medium access control protocol is proposed which enables the SU to merge (or aggregate) the sensed free bands for a single packet transmission without requiring the instantaneous tracking and estimation of the channel state information (CSI). The maximum stable throughput of a secondary node coexisting with multiple primary nodes is characterized and shown to be maximized by tuning the number of sensed subset of primary bands. We include sensing errors to the analysis and study the impact of the secondary number of antennas, $\mathcal{K}$, and the number of primary bands, $\mathcal{M}$, on the secondary throughput. To the best of our knowledge, the proposed protocol and its characterization from a network layer standpoint are addressed in this paper for the first time.

This paper is organized as follows. In the following Section, we present the system model adopted in this paper. The system analysis is discussed in Section \ref{secx}. In Section \ref{sec3}, we present some numerical results for the proposed protocol. We conclude the paper in Section \ref{sec4}.

%%%%% SYTEM MODEL
%%\vspace{-0.35cm}
\section{System Model}
\begin{figure}
  \centering
  \includegraphics[width=0.89\columnwidth]{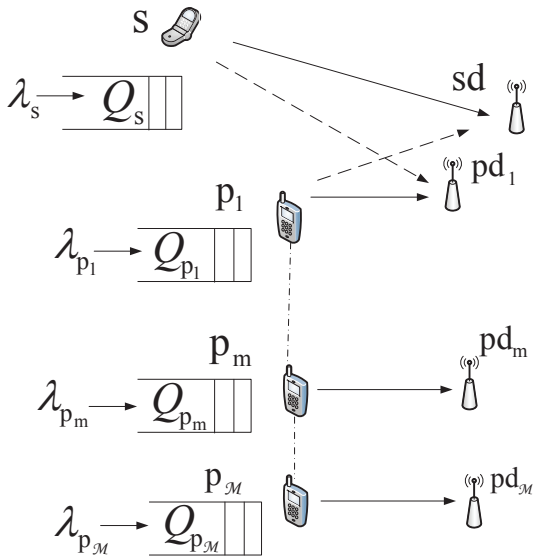}\\
   \caption{Primary and secondary links and queues. The solid lines denote the communication channels, whereas the dashed lines denote the interference channels. For clarity of the figure, we plotted the interference channels for the SU and user ${\rm p_1}$ only.}\label{fig00}
\end{figure}

%%\vspace{-0.24cm}
We consider a cognitive scenario with one SU equipped with $\mathcal{K}$ antennas and $\mathcal{M}$ PUs (as shown in Fig. \ref{fig00}). The PUs are communicating to their respective destinations using frequency division multiple-access technique. Specifically, we assume the existence of $\mathcal{M}$ PUs each of which is assigned to a unique orthogonal band. The ${\rm m}$th PU, ${\rm p_m}$, uses band ${\rm m}~\in~\{1,2,\dots,\mathcal{M}\}$ and wishes to communicate with its respective destination, ${\rm pd_m}$. The SU, ${\rm s}$, wishes to communicate with its respective destination, ${\rm sd}$. The channel is slotted and the length of one time slot is $T$. Each user has an infinite capacity buffer (queue) to store the incoming fixed-length packets, denoted by $Q_{\rm i}$ (see \cite{krikidis2010stability} for a similar assumption). The
arrivals at queue $Q_{\rm i}$ are independent and identically
distributed (i.i.d.) Bernoulli random variables \cite{krikidis2010stability} from slot to slot with mean $\lambda_{\rm i}\in[0,1]$ packets of size $b$ bits per time slot, `${\rm i}$' reads `${\rm p_m}$' for the queue of the PU assigned to band ${\rm m}$, and reads `${\rm s}$' for the secondary queue. Arrival
processes are independent from queue to queue and terminal to another \cite{sadek,krikidis2010stability}.

All wireless links exhibit fading and are corrupted by additive white Gaussian
noise (AWGN). The fading is assumed to be stationary, with
frequency non-selective Rayleigh block fading. This means
that the fading coefficient $h_{\rm i}$ (for the link connecting node ${\rm i}$ and its respective receiver) {\it remains
constant during one slot and over all frequency bands, but changes independently from one
slot to another} according to a circularly symmetric complex
Gaussian distribution with zero mean and variance $\sigma^2_{\rm i}$. Furthermore, the thermal noises at receivers are modeled as AWGN with zero mean and power spectral density $\mathcal{N}_\circ$ Watts/Hz. The CSI of links are known at the receivers only \cite{krikidis2010stability}. The primary node ${\rm p_m}$ has bandwidth ${ W_{\rm m}}\!=\!W$.

The cognitive radio user senses all bands simultaneously for $\tau$ seconds relative to the beginning of the time slot whose length is $T$ seconds. If the SU has $\mathcal{K}\le \mathcal{M}$ antennas, the time needed to sense $\mathcal{M}$ bands is $\tau=\lceil\mathcal{M}/\mathcal{K}\rceil \tau_B$, where $\lceil \cdot\rceil$ denotes the smallest integer greater than or equal to the argument and $\tau_B$ is the time spent in sensing one primary band.\footnote{If $\mathcal{K}> \mathcal{M}$, the time needed for sensing all bands is $\tau_B$, and the SU can assign some antennas to the same band to enhance the quality of the sensing process.} Since the cognitive radio user spends $\tau$ seconds in bands sensing, the remaining time for data transmission is $T\!-\!\tau$. An immediate observation is that as the sensing duration increases, the remaining time for data transmission decreases. Hence, the outage probability of the secondary link increases. All packets are assumed to be of the same length, and each contains $b$ bits. As will be explained later, the cognitive radio user merges the available bands. Therefore, if $\eta \leq \mathcal{M}$ bands are free, the transmission bandwidth is $\eta W$ with transmit power $\eta W \mathbb{P}_{\rm s}$ in Watts, where $\mathbb{P}_{\rm s}$ is the power spectral density of the SU in Watts/Hz.

The outage event occurs when the instantaneous capacity of the link is lower than the transmitted spectral efficiency rate.
Let $\sigma^2_{\rm p}$ denote the channel variance of user ${\rm p_m}$ and $\mathbb{P}_{\rm p}$ denote the transmit power spectral density of user ${\rm  p_m}$. The packet correct reception of the ${\rm m}$th PU is characterized by
the success probability \cite{krikidis2010stability,sadek}
\begin{equation}
\overline{P}_{\rm p_m}\!=\!{\rm Pr}\bigg\{\log_2\big(1 \!+\! \gamma_{\rm p}|h_{\rm p_m}|^2\big)\!>\!\mathcal{R}\bigg\}\!=\!\exp\bigg(-\mathcal{N}_\circ\frac{2^{\mathcal{R}}\!-\!1}{\sigma^2_{\rm p}\mathbb{P}_{\rm p}}\bigg),
\label{choutage}
\end{equation}
 where ${\rm Pr}\{\cdot\}$ denotes the probability of the event in the argument and $\mathcal{R}=b/(W T)$. We note that the probability of packet correct reception (complement of channel outage) increases with $\sigma^2_{\rm p}\mathbb{P}_{\rm p}$. Moreover, it decreases with $\mathcal{R}$ and $\mathcal{N}_\circ$.

 For simplicity of presentation, we assume symmetric PUs, which implies that all PUs have the same arrival rate and the same channel parameters \cite{bao2010stable}. Hence, $\lambda_{\rm p_m}\!=\!\lambda_{\rm p}$, $\overline{P}_{\rm p_m}\!=\!\overline{P}_{\rm p}$, and the sensing errors probabilities are equal for all primary bands. The essential difference is that for the general
asymmetric case, the analysis has to keep track of the different
possible primary transmitters and channel sets, which clutters the
notation while the approach remains similar to the symmetric case.

 We assume that the SU uses the antenna with the highest average channel gain (variance of complex fading coefficient) to its destination for packets transmission. Using such antenna will provide the highest correct packet reception for the secondary packets. This can be seen from the correct reception probability in (\ref{choutage2}) and (\ref{choutage}) where the correct packet reception is monotonically decreasing with the variance of the channel between the transmitter of the packet and its intended receiver. We denote the highest expected channel gain between the SU and its respective receiver by $\sigma^2_{\rm s}$. Assume that the SU detects $\eta$ bands to be free, if all these bands are truly free of any primary transmissions, the probability of correct packet reception of the SU is given by
\begin{equation}
\overline{P}_{\rm s,\eta W}\!=\!\exp\bigg(-\mathcal{N}_\circ\frac{2^{\mathcal{R}_{\eta}}\!-\!1}{\sigma^2_{\rm s}\mathbb{P}_{\rm s}}\bigg),
\label{choutage2}
\end{equation}
with
 \begin{equation}
\mathcal{R}_{\eta}\!=\!\frac{b}{\eta W(T\!-\!\lceil\!\frac{\mathcal{M}}{\mathcal{K}}\!\rceil\tau_B)^+}\!
%=\!\frac{b}{\eta WT(1\!-\!\lceil\!\frac{\mathcal{M}}{\mathcal{K}}\!\rceil\frac{\tau_B}{T})^+}\!
=\!\frac{\mathcal{R}}{\eta (1\!-\!\lceil\!\frac{\mathcal{M}}{\mathcal{K}}\!\rceil\frac{\tau_B}{T})^+}
\end{equation}
 where $(V)^+$ denotes $\max\{V,0\}$. If there are two concurrent transmissions over any band, both packets under transmission will be lost. Increasing the number of antennas allows the SU to invest more time in data transmission, hence increases the secondary throughput. Whereas, the huge increase of the number of primary bands may cause time slot consumption, hence successful transmission with probability zero.
 We assume that the cognitive radio user cannot send more than one packet at any slot. The medium access control operation can be described as follows.
\begin{itemize}
\item The PUs access the channel at the beginning of the time slot if their queues are nonempty.
\item The SU senses all the primary bands from the beginning of the time slot to $\tau$ seconds to detect the possible activity of the PUs.
\item The SU merges (aggregates) the available bandwidths of the sensed free bands, and sends exactly one packet.
\item At the end of each time slot, a feedback signal from the respective receiver is sent to inform the transmitter about the status of its packet decodabiliy\footnote{As usual, we assume that errors in the feedback messages are negligible, which is reasonable for short length packets as strong and low rate
codes can be employed in the feedback channel \cite{krikidis2010stability,sadek}.}. If the intended receiver can decode the packet, it sends back an acknowledgement (ACK) message; otherwise it sends back a negative-acknowledgement (NACK) message.
\item A correctly received packet is removed from the respective transmitter's queue.
\item In the case of packets loss due to concurrent transmission or channel impairments, re-transmission of the lost data is required.
\end{itemize}
%%\vspace{-0.4cm}
\section{System Analysis}\label{secx}
%%\vspace{-0.2cm}
%%\vspace{-0.2cm}
A fundamental performance measure of a communication network is the stability of the queues. Stability can be defined rigorously as follows: Denote by $Q^{\left(t\right)}$ the length of queue $Q$ at the beginning of time slot $t$. Queue $Q$ is said to be stable if \cite{sadek} $\lim_{z \rightarrow \infty  } \lim_{t \rightarrow \infty  } {\rm Pr}\{Q^{\left(t\right)}<z\}=1$. A system is said to be stable, if every queue belonging to it is stable. Loynes' theorem \cite{sadek} states that if the arrival process and the service process of a queue are strictly stationary, then the queue is stable if and only if the average service rate is greater than the average arrival rate of the queue.

Let $X^t_{\rm i}$ denote the number of arrivals to $Q_{\rm i}$ in an arbitrary time slot $t$, and $Y^t_{\rm i}$ denote the number of departures from $Q_{\rm i}$ in an arbitrary time slot $t$. Based on the late arrival model, which means that an arriving packet will be blocked of service during its arrival time slot even if the queue is empty, the evolution of $Q_{\rm i}$ is given by
\begin{equation}
\begin{split}
Q_{\rm i}^{t+1}=(Q_{\rm i}^t-Y_{\rm i}^t)^++X^t_{\rm i}
\end{split}
\end{equation}

Due to the presence of the sensing errors, the SU may interfere with the PUs and therefore packet collision and throughput loss may occur. As will be discussed later, the SU will suffer from throughput degradation due to sensing errors. Let $P_{\rm FA}$ denote the probability that the SU's sensor generates false alarms, and $P_{\rm MD}$ denote the probability that the SU misdetects the primary activity.\footnote{Since the decision on the activity of each of the PUs is taken separately, i.e., the SU senses and decides on the basis of the sensing outcome of each band independently, the probability of misdetecting the ${\rm m}$th PU is $P_{\rm MD}$ regardless of other PUs' state. We note that the sensing algorithm quality depends on the sensing duration, $\tau_B$, the channel between the PU and the SU, the primary transmit powers, and many other parameters.}

Since the queues service rates are coupled, i.e., interacting queues, we resort to the approach of dominant systems (or saturated/backlogged SU) \cite{sadek,rao1988stability}. In a dominant system, the behaviour of nodes and channels realization are the same, but the dependent node (secondary node), if emptied, sends dummy packets. These dummy packets may interfere with the PUs, but do not contribute to the SU throughput. This idea has been used in many works (e.g. \cite{rao1988stability,sadek,krikidis2010stability} and the references therein). It has been shown that the stability of the dominant system and the original system are indistinguishable at the boundary points. Hence, the original system is stable if and only if the dominant system is stable.

\subsection{Stability Region Analysis Using The Dominant System}

Since the SU is saturated (always backlogged), the probability of ${\rm m}$th PU successfully transmits its data packet is given by the event that the link between ${\rm p_m}$ and its respective receiver is not in outage and that the SU detects the primary active correctly. Hence, the mean service rate of the ${\rm m}$th PU is
\begin{equation}
\begin{split}
\mu_{\rm p_m}&=\mu_{\rm p}=\overline{P}_{\rm  p} (1-P_{\rm MD})
\end{split}
\end{equation}

Let $\pi\!=\!1\!-\!\lambda_{\rm p}/\mu_{\rm p}$ denote the probability of the ${\rm m}$th PU being empty \cite{sadek,bao2010stable,krikidis2010stability}. When $\eta$ primary bands are free, the SU must detect at least one of them to be free correctly in order for service to occur. Also, it must detect the activity of all the active users correctly to avoid concurrent transmission and packets loss. Hence, the mean service rate of the secondary queue is given by
\begin{equation}
\begin{split}
\mu_{\rm s}&\!=\!\sum_{\eta\!=\!1}^{\mathcal{M}} \Biggr(\!\begin{array}{c}
               \mathcal{M} \\
               \eta
             \end{array}\!\Biggr)
\pi_{\rm }^\eta \overline{\pi}_{\rm }^{(\mathcal{M}\!-\!\eta)}  \overline{P}_{\rm MD}^{(\mathcal{M}\!-\!\eta)}  \sum_{n=1}^{\eta} \Biggr(\!\begin{array}{c}
               \eta \\
               n
             \end{array}\!\Biggr)\overline{P}_{\rm FA}^n P_{\rm FA}^{(\eta-n)} \overline{P}_{{\rm  s},n W}
\end{split}
\end{equation}
where $\Big(\!\begin{array}{c}
               y \\
               x
             \end{array}\!\Big)$ denotes $y$ choose $x$ and the term $\overline{P}_{\rm MD}^{(\mathcal{M}\!-\!\eta)}$ means that the SU must not use the band of an active PU, which will cause packet loss for the SU and the PU; and the term $(\begin{array}{c}
               \eta \\
               n
             \end{array})\overline{P}_{\rm FA}^n P_{\rm FA}^{(\eta-n)}$ represents the probability of generating false alarm over $\eta-n$ bands out of the $\eta$ empty bands.

Based
on the construction of the dominant system, the data queues
of the dominant system are always larger in size than those of
the original system, provided that the queues start with identical
initial conditions in both systems. Therefore, for a given $\lambda_{\rm p}\!\le\! \overline{P}_{\rm  p} (1-P_{\rm MD})$, if for some $\lambda_{\rm s}$,
the queue $Q_{\rm s}$ is stable in the dominant system then it must
be stable also in the original system. Conversely, if for some $\lambda_{\rm s}$ in
the dominant system, the queue $Q_{\rm s}$ saturates, then it will send a physical packet instead of transmitting dummy packets and thus the behavior of the dominant system becomes identical to that of the original system.
Therefore, the original system and the dominant system are
{\bf indistinguishable} at the boundary points and thus have the
same stability region.

Using Loynes' theorem, the stability region, $\mathcal{S}$, is given by
   \begin{equation}
\begin{split}
   &\mathcal{S}\!=\!\Biggr\{(\lambda_{\rm p},\lambda_{\rm s}):\lambda_{\rm s} \!< \!\sum_{\eta=1}^{\mathcal{M}} \Bigg[\Biggr(\!\begin{array}{c}
               \mathcal{M} \\
               \eta
             \end{array}\!\Biggr)
\pi_{\rm }^\eta \overline{\pi}_{\rm }^{(\mathcal{M}\!-\!\eta)}  \overline{P}_{\rm MD}^{(\mathcal{M}\!-\!\eta)} \\& \,\,\,\,\,\,\,\,\,\,\,\,\,\,\,\,\,\,\,\,\,\,\,\,\,\,\,\,\,\,\,\,\,\,\,\,\,\,\,\,\,\,\,\,\,\,\,\,\,\,\,\,\,\,\,\,\,\,\,\,\,\   \times \sum_{n=1}^{\eta} \Biggr(\!\begin{array}{c}
               \eta \\
               n
             \end{array}\!\Biggr)\overline{P}_{\rm FA}^n P_{\rm FA}^{(\eta-n)} \overline{P}_{{\rm  s},n W} \Bigg]\!\Biggr\}
\end{split}
\end{equation}
with $\lambda_{\rm p}<\mu_{\rm p}$.
%%\vspace{-0.4cm}
\subsection{Optimal Number of Sensed Bands}
%%\vspace{-0.2cm}
As the number of PUs increases, the secondary throughput increases due to the possibility of exploiting more free bands. However, due to sensing errors, for specific value of $\mathcal{M}$, the secondary throughput will start to degrade as the number of PUs increases, due to the increase in the probability of misdetecting one of the active PUs, as will be shown in the numerical results. One can think about finding the optimal number of sensed bands, $\mathcal{M}_\circ\!\le\! \mathcal{M}$, which should be selected by the SU from the available primary bands such that its throughput is maximized under the stability of the primary queues.
Given that all the PUs are stable, i.e., $\pi>0$, the optimization problem that describes the optimal secondary maximum stable throughput for each $\lambda_{\rm p}$ can be stated as follows:
 \begin{equation}
\begin{split}
\underset{\mathcal{M}_\circ\le \mathcal{M}}{\max.} \,\,\ \! & \mu_{\rm s}, \ \pi\!\ge\!0
\end{split}
\end{equation}
The optimization problem can be solved via a simple grid search over the integer values from $\mathcal{M}_\circ=1$ to $\mathcal{M}_\circ=\mathcal{M}$.

 It should be noted that if the SU is {\it a power-limited device}, i.e., if the transmit power per time slot is constrained by a certain upper (maximum) value, i.e., $ \mathbb{P}_\circ= W \mathbb{P}_{\rm s}$, where $\mathbb{P}_\circ$ is the total allowable power per time slot in Watts; the sole variation in the above analysis is in the value of the complement of channel outage. Specifically, when there are $\eta \leq \mathcal{M}$ free bands, and the SU sensed them to be inactive, the probability of secondary packet correct reception is given by
 \begin{equation}
\overline{P}_{\rm s,\eta W}\!=\!\exp\bigg(-\eta \mathcal{N}_\circ\frac{2^{\mathcal{R}_{\eta}}\!-\!1}{\sigma^2_{\rm s}\mathbb{P}_{\rm s}}\bigg),
\label{choutage}
\end{equation}

For multiple SUs with a single destination, we can assume that the SUs share the spectrum using random multi-access channel system \cite{krikidis2010stability} (where users can simultaneously use the spectrum). For general systems with multiple secondary destinations, the SUs can adopt time division multiple-access schemes. Moreover, the SUs can cooperatively detect the available bands\footnote{Which could enhance the quality of channel sensing probabilities as an
appropriate fusion and combination of the individual sensing
data improves the ability of the system and decreases the
probability of detection error (misdetction) and false alarm.} and split them such that each SU obtains an orthogonal subset of the sensed free bands for the transmission of its packet.

%%\vspace{-0.3cm}
    \section{Numerical Results}\label{sec3}
%%\vspace{-0.2cm}

    In this section, we present some numerical results for the presented optimization problems in this paper. %Figs. \ref{fig1}-\ref{fig4} and \ref{fig7} are for the unlimited power case.
    \begin{figure}
  \centering
  \includegraphics[width=0.88\columnwidth]{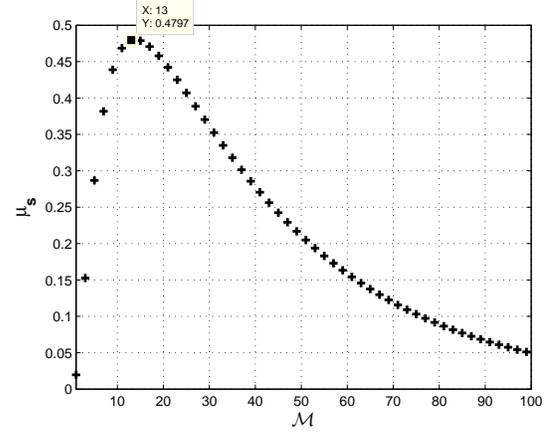}\\
   \caption{Maximum mean secondary service rate versus the number of PUs. The parameters used to generate the figure are: $\sigma^2_{\rm s} {\mathbb{P}_{\rm s}}/{\mathcal{N}_\circ}=1$, $\mathcal{K}=8$, $\tau_B=0.01T$, $\mathcal{R}=b/(TW)=2$ bits/sec/Hz, $\overline{P}_{\rm  p}=0.9$, $P_{\rm MD}=0.05$, $P_{\rm FA}=0.05$, and $\lambda_{\rm p}=0.5$ packets per time slot.}\label{fig1}
   %%\vspace{-0.2cm}
\end{figure}

    \begin{figure}
  \centering
  \includegraphics[width=0.89\columnwidth]{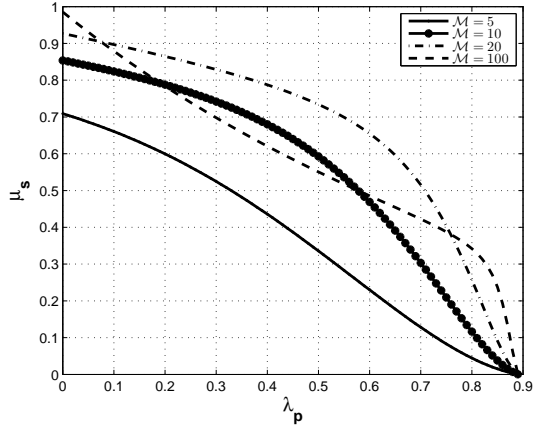}\\
   \caption{Maximum mean secondary service rate for the proposed protocol versus $\lambda_{\rm p}$. The parameters used to generate the figure are: $\sigma^2_{\rm s} {\mathbb{P}_{\rm s}}/{\mathcal{N}_\circ}=1$, $\mathcal{K}=8$, $\tau_B=0.01T$, $\mathcal{R}=b/(TW)=2$ bits/sec/Hz, $\overline{P}_{\rm  p}=0.9$, $P_{\rm MD}=0.01$, and $P_{\rm FA}=0.05$.}\label{fig2}
     \vspace{-0.3cm}
\end{figure}

  \begin{figure}[t]
  \centering
  \includegraphics[width=0.89\columnwidth]{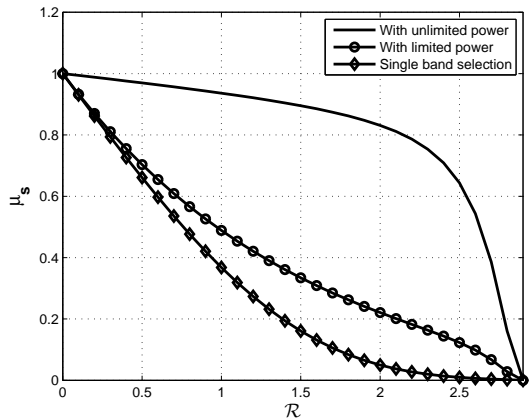}\\
   \caption{Maximum throughput of the proposed protocol with and without power constrained vs. a single band transmit system. The parameters used to generate the figure are: $\sigma^2_{\rm s} {\mathbb{P}_{\rm s}}/{\mathcal{N}_\circ}=1$, $\mathcal{K}=8$, $\tau_B=0.01T$, $\mathcal{M}=15$, $P_{\rm MD}=P_{\rm FA}=0$, $\lambda_{\rm p}=0.2$ packets/slot, and $\sigma^2_{\rm p} {\mathbb{P}_{\rm p}}/{\mathcal{N}_\circ}=4$.}\label{fig6}
   \vspace{-0.3cm}
\end{figure}

  \begin{figure}
  \centering
  \includegraphics[width=0.89\columnwidth]{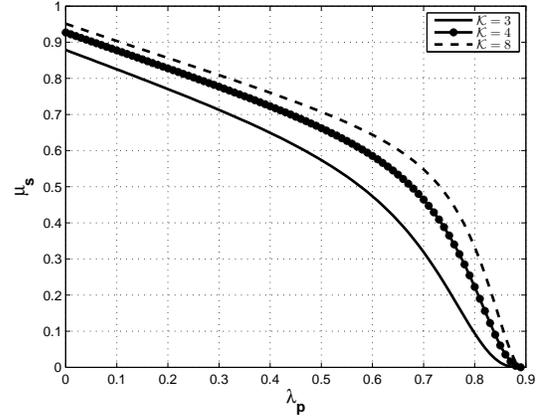}\\
   \caption{The impact of SU's number of transmit antennas on stability. The parameters used to generate the figure are: $\sigma^2_{\rm s} {\mathbb{P}_{\rm s}}/{\mathcal{N}_\circ}=1$, $\mathcal{K}=8$, $\mathcal{R}=b/(TW)=2$ bits/sec/Hz, $\overline{P}_{\rm  p}=0.9$, $P_{\rm MD}=0.01$, $P_{\rm FA}=0.05$, $\tau_B=0.05 T$, and $\mathcal{M}=40$.}\label{fig7}
     \vspace{-0.4cm}
\end{figure}

    In Fig. \ref{fig1}, we plot the secondary throughput versus $\mathcal{M}$. The parameters used to generate the figure are: $\sigma^2_{\rm s} {\mathbb{P}_{\rm s}}/{\mathcal{N}_\circ}=1$, $\mathcal{K}=8$, $\tau_B=0.01T$, $\mathcal{R}=b/(TW)=2$ bits/sec/Hz, $\overline{P}_{\rm  p}=0.9$, $P_{\rm MD}=0.05$, $P_{\rm FA}=0.05$, and $\lambda_{\rm p}=0.5$ packets per time slot. From the figure, the secondary throughput increases with $\mathcal{M}$ until specific $\mathcal{M}\!=\!\mathcal{M}_\circ$ where the behavior is reversed. It is noted that if the SU chooses only $\mathcal{M}_\circ\!=\!13$ bands to exploit when some or all of them become empty, the secondary stable throughput is maximized.

Fig. \ref{fig2} shows the maximum secondary stable throughout of the proposed system for different values of $\mathcal{M}$. The parameters used to generate the figure are: $\sigma^2_{\rm s} {\mathbb{P}_{\rm s}}/{\mathcal{N}_\circ}=1$, $\mathcal{K}=8$, $\tau_B=0.01T$, $\mathcal{R}=b/(TW)=2$ bits/sec/Hz, $\overline{P}_{\rm  p}=0.9$, $P_{\rm MD}=0.01$, and $P_{\rm FA}=0.05$. As shown in the figure, the secondary stable throughout for each $\lambda_{\rm p}$ expands as the number of the primary bands increases up to a specific $\mathcal{M}$. For $\mathcal{M}=100$, $\mu_{\rm s}$ for some $\lambda_{\rm p}$ is higher than $\mu_{\rm s}$ for the lower $\mathcal{M}$ curves and for other values the behavior is reversed.

    Fig. \ref{fig6} shows the maximum secondary stable throughput for the system with and without maximum power constraint per time slot on the secondary transmit power. For comparison purposes, the system in which the SU selects one of the empty bands, when there is more than one band free, and transmits with its full power is presented. The parameters used to generate the figure are: $\sigma^2_{\rm s} {\mathbb{P}_{\rm s}}/{\mathcal{N}_\circ}=1$, $\mathcal{K}=8$, $\tau_B=0.01T$, $\mathcal{M}=15$, $P_{\rm MD}=P_{\rm FA}=0$, $\lambda_{\rm p}=0.2$ packets per time slot, and $\sigma^2_{\rm p} {\mathbb{P}_{\rm p}}/{\mathcal{N}_\circ}=4$. The figure reveals the supremacy of our proposed protocol (under limited and unlimited power assumptions) over the single band selection protocol.

    The impact of the SU's number of antennas on the stability region is shown in Fig. \ref{fig7}. The parameters used to generate the figure are: $\sigma^2_{\rm s} {\mathbb{P}_{\rm s}}/{\mathcal{N}_\circ}=1$, $\mathcal{K}=8$, $\mathcal{R}=b/(TW)=2$ bits/sec/Hz, $\overline{P}_{\rm  p}=0.9$, $P_{\rm MD}=0.01$, $P_{\rm FA}=0.05$, $\tau_B=0.05 T$, and $\mathcal{M}=40$. The fact that the throughput increases with increasing the number of antennas is shown in the figure. However, this increase is limited by the factor $\lceil\!{\mathcal{M}}/{\mathcal{K}}\!\rceil{\tau_B}/{T}$. If this factor is negligible, i.e., $\lceil\!{\mathcal{M}}/{\mathcal{K}}\!\rceil{\tau_B}/{T}\approx 0$, increasing $\mathcal{K}$ will have almost no effect on the secondary throughput.

\vspace{-0.1cm}
	\section{Conclusions}\label{sec4}
\vspace{-0.1cm}
%%\vspace{-0.2cm}
   We have proposed a novel access scheme for cognitive radio users. The cognitive radio user combines (merges) the available primary bands to increase the probability of successful packet reception of its data packets, which consequently increases its service rate. Our results have revealed that there is an optimal value for the number of sensed primary bands that maximizes the secondary throughput.
   A possible extension of the results presented in this work is to consider a multi-packet reception channel model which allows concurrent transmissions. In this case, packets can survive from interference if the received signal-to-noise-and-interference-ratio (SINR) is greater than a certain decoding threshold.
   \vspace{-0.1cm}
   \section{Acknowledgement}
   \vspace{-0.1cm}
   This paper was made possible by a NPRP grant 6-1326-2-532 from the
Qatar National Research Fund (a member of The Qatar Foundation). The
statements made herein are solely the responsibility of the authors.
%%\vspace{-0.4cm}
\bibliographystyle{IEEEtran}
\bibliography{IEEEabrv,term_bib}
\end{document}